\documentclass[a4paper,11pt]{article}
\usepackage{jinstpub} 
\usepackage{subfigure,dcolumn}
\usepackage{epstopdf}
\usepackage[version=4]{mhchem}
\usepackage{upgreek}
\usepackage{graphicx}
\usepackage{xcolor}
\usepackage{amsmath}
\usepackage{ulem}
\usepackage{multirow}
\usepackage{caption}

\title{\boldmath Non-cyclic prototype of JUNO-TAO VETO water tank with 3 inch PMTs}

\author[a]{Ziming Li,}
\author[b,c,d]{Zhimin Wang,}
\author[a]{Jie Yang,}
\author[b,c]{Min Li,}
\author[b]{Diru Wu,}
\author[b,c]{Yichen Zheng,}
\author[b]{Wei He,}
\author[b,c]{Ruhui Li,}
\author[b,c]{Hongjian Li,}
\author[b,c]{Derun Wang,}
\author[b,c]{Ziliang Chu,}
\author[b,c]{Miao He,}
\author[b,c]{Jilei Xu,}
\author[b]{Yangfu Wang,}
\author[b,c]{Shuihan Zhang,}
\author[b,c]{Xiaolu Ji,}
\author[e]{Yongbo Huang,}
\author[f]{Abdel Rebii,}
\author[f]{Cedric Cerna,}
\author[f]{Frédéric Druillole,}
\author[f]{Frédéric Perrot,}
\author[g]{Loïc  Labit,}
\author[h]{Mariangela  Settimo,}
\author[i]{Juan Pedro Ochoa-Ricoux,}
\author[j,k]{Angel Abusleme,}
\author[j,k]{Pablo Walker,}
\author[j,k]{Agustin Campeny,}
\author[j]{Rafael Herrera,}
\author[j]{Giancarlo Troni,}
\author[j]{Ignacio Jeria}

\affiliation[a]{Zhengzhou University, Henan 450001, China}
\affiliation[b]{Institute of High Energy physics, Beijing 100049, China}
\affiliation[b,c]{University of Chinese Academy of Sciences, Beijing 100049, China}
\affiliation[d]{State Key Laboratory of Particle Detection and Electronics, Beijing 100049, China}
\affiliation[e]{Guangxi University, Guangxi 530005, China}
\affiliation[f]{University of Bordeaux, CNRS, LP2i, UMR 5797, F-33170 Gradignan, France}
\affiliation[g]{SUBATECH, CNRS/IN2P3, Université de Nantes, IMT Atlantique, Nantes, France}
\affiliation[h]{Institut Pluridisciplinaire Hubert Curien, Université de Strasbourg, F-67037 Strasbourg, France}
\affiliation[i]{Department of Physics and Astronomy, University of California, Irvine, Irvine, California, USA}
\affiliation[j]{Pontificia Universidad Católica de Chile, Santiago, Chile}
\affiliation[k]{Millennium Institute for SubAtomic Physics Physics at the High-energy Frontier (SAPHIR), ANID, Chile}

\emailAdd{wangzhm@ihep.ac.cn}
\emailAdd{yangjie@zzu.edu.cn}

\abstract{
The Taishan Antineutrino Observatory (TAO, also known as JUNO-TAO) is a satellite experiment of the Jiangmen Underground Neutrino Observatory (JUNO). A ton-level liquid scintillator detector will be placed at about 30\,m from a core of the Taishan Nuclear Power Plant. The reactor antineutrino spectrum will be measured with sub-percent energy resolution, to provide a reference spectrum for future reactor neutrino experiments, and to provide a benchmark measurement to test nuclear databases. 
A Cerenkov water tank system with a thickness of 1.2\,m pure water will be located around the central detector of TAO. The water tank system designed with 300 3" PMTs (SPMT) will use the same electronics as JUNO SPMT, but with online software multiplicity trigger. The performance of a Cerenkov detector with the JUNO SPMT and electronics designed for liquid scintillator detector needs to be checked, including the software triggering. The features and the long stability of the detector without water circulation also needs to be checked as a common concern and a backup option for future JUNO-TAO running.
Here we will summary the integration and testing of a prototype water tank detector system, including SPMT, electronics, data taking, simulation and measurement results.
}


\arxivnumber{1234.56789} 

\begin{document}
\maketitle
\flushbottom

\section{Introduction}
\label{1:intro}

The Taishan Antineutrino Observatory (TAO or JUNO-TAO)\,\cite{2020arXiv200508745J,Steiger2022TAOTheTA} is a satellite detector for the Jiangmen Underground Neutrino Observatory (JUNO)\,\cite{JUNO-2022103927}. 
TAO designed with a ton-level liquid scintillator detector located at about 30\,m from a core of the Taishan Nuclear Power Plant, and will realize a reactor neutrino detection rate of about 2000 per day.
It will measure the reactor antineutrino spectrum with high precision and high energy resolution to provide a reference spectrum for JUNO and other reactor antineutrino experiments, and provide a benchmark measurement to test nuclear databases.

Short-baseline reactor antineutrino experiments with shallow overburden usually have large cosmogenic neutron backgrounds, which also is one of the critical concerns of TAO since the overburden is just 10 meter-water-equivalent. 
In order to reduce the neutron and shield radioactivity backgrounds, a muon veto detector system was designed with optimization of the veto strategy\,\cite{Li_2022,li2022ambientneutronmeasurementtaishan}, including a dodecagon water tank, a polyethylene layer above the bottom lead shield, and a top veto tracker\,\cite{limin-compact-plastic,Luo2023DesignOO} above the top polyethylene shielding layer of TAO central detector.The PMT coverage of the JUNO-TAO VETO TANK is approximately 1\%, and the future requirement for muon detection efficiency is greater than 95\%.

The quality of water is a key factor for a Cerenkov detector performance, in particular for a non-cyclic tank running under an uncontrolled environment in a long term\,\cite{auger-prototype-ABRAHAM200450,auger-2015172, water-prototype-dyb-YU201226}. The electronics of JUNO small photomultiplier tube (SPMT)\,\cite{catiroc-10,Blin_2017,JUNO-double-miao} is designed for liquid scintillator (LS) detector, which has a very different light intensity and photon hit time distribution to a water Cerenkov detector. A software trigger is designed by TAO VETO water tank system, which will be a different configuration to what JUNO planned. It is even worse when the used 3-inch PMTs has a higher dark count rate (DCR) than that of used by JUNO. To check and answer all the concerns, a 1\,m$^3$ water tank with 16 3-inch PMTs is designed and realized including the electronics. 

In this paper, the design of the prototype detector including the 3-inch PMTs, electronics and DAQ, and the construction will be discussed in Sec.\,\ref{1:prototype}. The onsite PMT calibration and the event assembly will discussed in Sec.\,\ref{1:exp}. Then the key features and stability of the prototype will be shown in Sec.\,\ref{1:results}. 
A preliminary simulation of the detector and a quick comparison with data will be discussed in Sec.\,\ref{1:sim}. Finally a short summary is reached in Sec.\,\ref{1:summary}.

\section{Prototype of water tank}
\label{1:prototype}

\subsection{3-inch PMT}

25,000 3-inch photomultiplier tubes (SPMTs)\,\cite{linan2021165347-3inch,CAO2021165347-3inch,diru2021165347-3inch} had been selected for Jiangmen Underground Neutrino Observatory (JUNO) produced by the Hainan Zhanchuang Photonics Technology Co., Ltd (HZC) company in China. Fifteen performance parameters were tracked at different sampling rates. 300 and 16 SPMTs are further ordered in two batches from HZC which were the rejected tubes by JUNO, and plan to install them in JUNO TAO VETO water tank detector. 16 of the tubes are applied firstly in the small water tank prototype.

\begin{figure}
    \centering
    \includegraphics[width=1\linewidth]{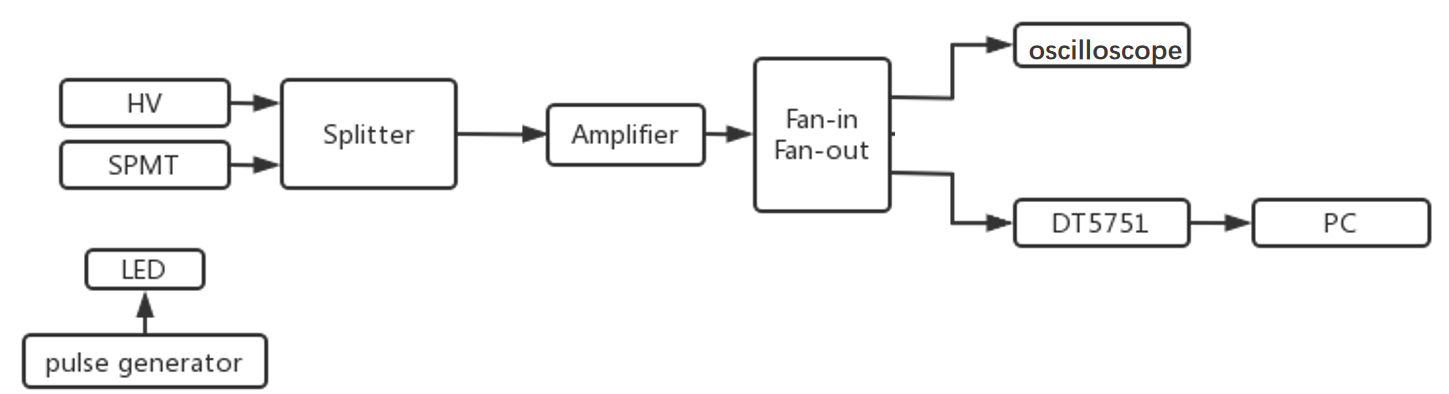}
    \caption{Logic flowchart of PMT calibration}
    \label{fig.Logic Flowchart}
\end{figure}

\begin{figure}[!ht]
\centering
\subfigure[Waveform]{\includegraphics[width=0.6\textwidth]{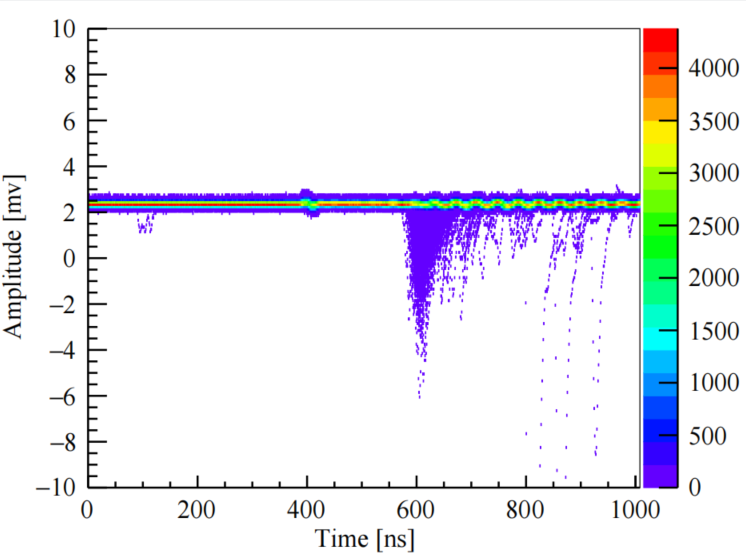}}
\subfigure[SPE Charge vs.\, HV]{\includegraphics[width=0.46\textwidth]{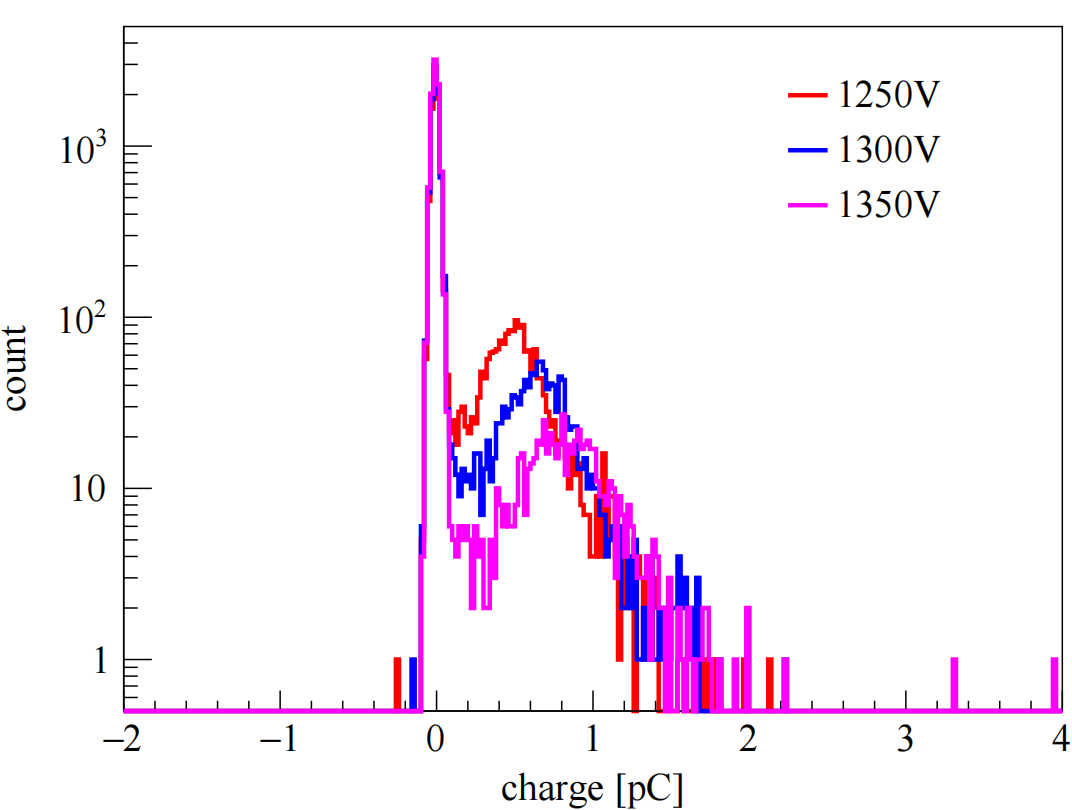}}
\subfigure[Fitting of Gain vs.\, HV]{\includegraphics[width=0.47\textwidth]{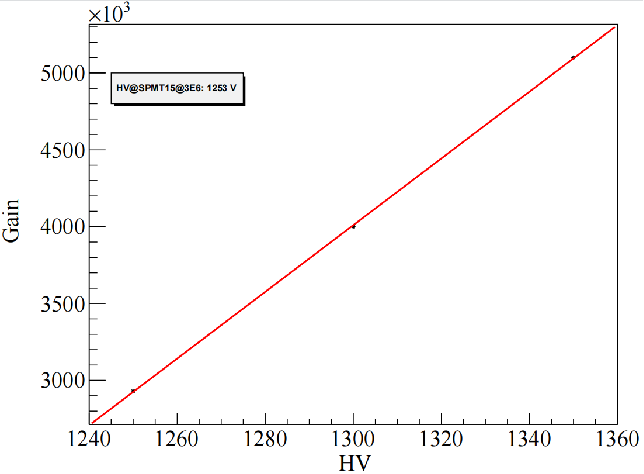}}
\caption{ SPMT testing}
\label{fig.d}
\end{figure}

\begin{figure}
    \centering
    \includegraphics[width=0.8\linewidth]{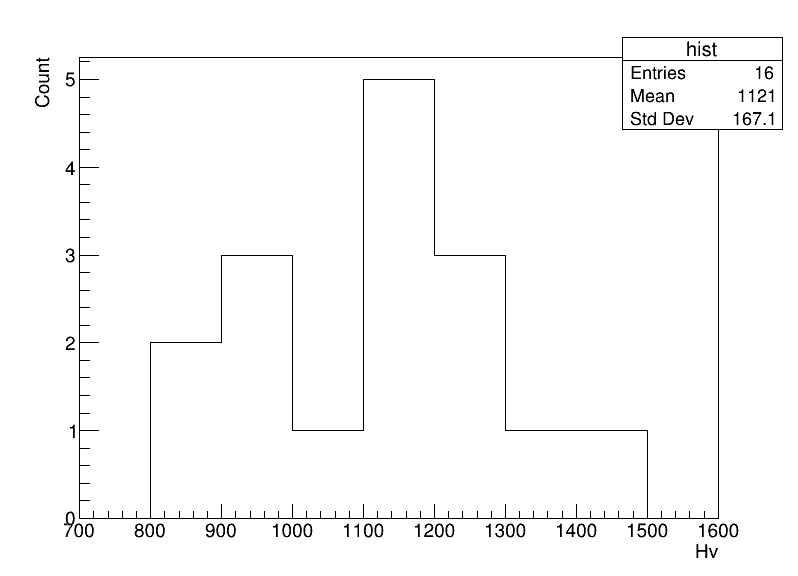}
    \caption{Distribution of nominal voltage of PMTs.}
    \label{fig.nominal HV}
\end{figure}

The 16 3-inch PMTs, before installation, are further measured one by one in a dark box 
by a testing system (Fig.\,\ref{fig.Logic Flowchart}) with an LED illumination under single photon (SPE) level. DT5751\,\cite{DT5751} is the key of the testing system, which is helping to collect the PMT output waveforms
, calculate the output charge, 
and obtain the gain versus high voltage (HV) by surveying several voltages as shown on Fig.~\ref{fig.d}.
Finally, we tested all the 16 3-inch PMTs by single photon electron (p.e.) and the LED illumination to find the aimed voltage for a gain of $3\times10^6$.
The nominal voltage for all of the PMTs is as shown in Fig.~\ref{fig.nominal HV}. It can be seen that the average voltage is 1120\,V. To prevent some PMTs from operating at excessively low voltages, which could affect detection efficiency, the operating voltage has been set to 1200\,V. There is a significant difference among the 16 PMTs if we considering the working mode of the used electronics, where 16 of the PMTs connected to the same group only can be supplied with a same HV.
At the same time, the rise-time, fall-time, and FWHM of the 16 PMTs were obtained too, as shown in Fig.~\ref{fig.rise time, fall time, and FWHM}. The average rise-time is around 7\,ns, the average of fall-time is around 17\,ns, and the average of FWHM is around 13\,ns.




\begin{figure}[!ht]
    \centering
    \includegraphics[width=0.8\linewidth]{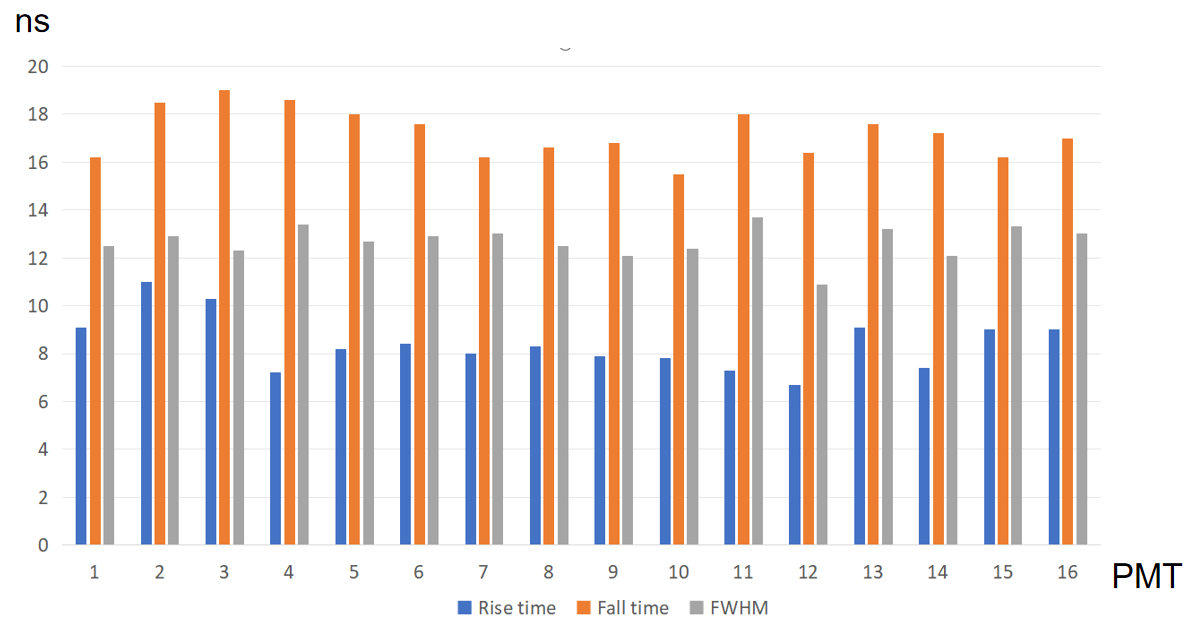}
    \caption{The rise-time, fall-time, and FWHM of 16 PMTs}
    \label{fig.rise time, fall time, and FWHM}
\end{figure}

\subsection{Electronics and DAQ}

JUNO (Jiangmen Underground Neutrino Observatory) experiment, a liquid scintillator antineutrino detector with a double calorimetry system, combines about 17k 20-inch PMTs (Large PMTs system)\,\cite{JUNO20inchPMT} and around 25k 3-inch PMTs (Small PMTs system)\,\cite{juno-double-calorimetry-miao,JUNO3inchPMT}. The ASIC CATIROC (Charge And Time Integrated Read Out Chip) as a complete read-out chip designed to read arrays of 16 photomultipliers (PMTs), is applied to the 3-inch PMTs system. CATIROC is a SoC (System on Chip) that processes analog signals up to the digitization to reduce the cost and cables number\,\cite{10.1007/978-981-13-1313-4_34}. The ASIC is composed of 16 independent channels that work in trigger less mode, auto-triggering on the single photo-electron (PE). It provides a charge measurement with a charge resolution of 15\,fC and a timing information with a precision of 200\,ps rms\,\cite{Signal-SPMT-diru}. 

\begin{figure}[!ht]
    \centering
    \includegraphics[width=0.5\linewidth]{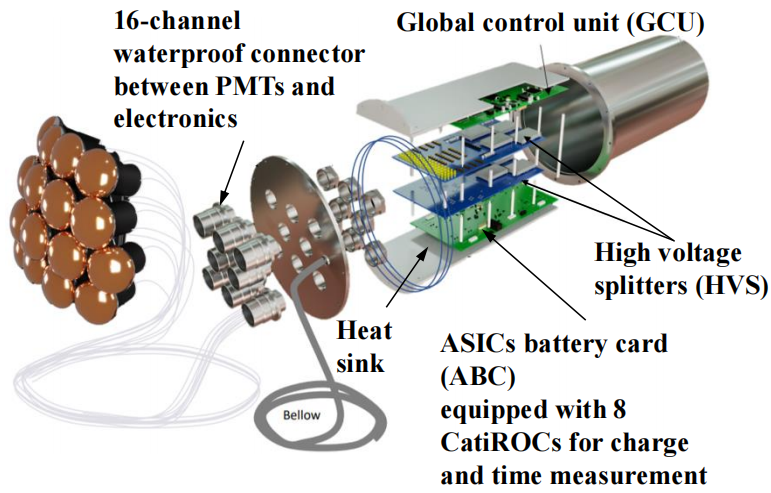}
    \caption{JUNO under water box (UWB)\cite{JUNO-2022103927}}
    \label{fig:enter-label}
\end{figure}

JUNO-TAO adopts a similar design with SPMTs for its VETO water tank. Here, we test the small water tank prototype with only 16 PMTs by an ASIC CATIROC of a JUNO underwater box (UWB) (Fig.\,\ref{fig:enter-label}), which contains in total 128 individual channels. While only 16 individual channels are turned on with the same HV value. The data of each individual channel will be taken by the triggerless and auto-triggering mode according to its own settable threshold on amplitude of around 0.3\,p.e. The record data includes the integrated charge in 10 bits high gain to cover 0-10\,p.e.\,for $3\times10^{6}$ gain and a coarse gain in 10 bits to cover 10-100\,p.e., and a global time stamp including a coarse range of 26 bits in a unit of 25\,ns and a fine range of 10 bits in unit of 25/1024\,ns.

\begin{figure}[!ht]
    \centering
    \subfigure[Pedestal of single channel in high and coarse range]{\includegraphics[width=0.48\textwidth]{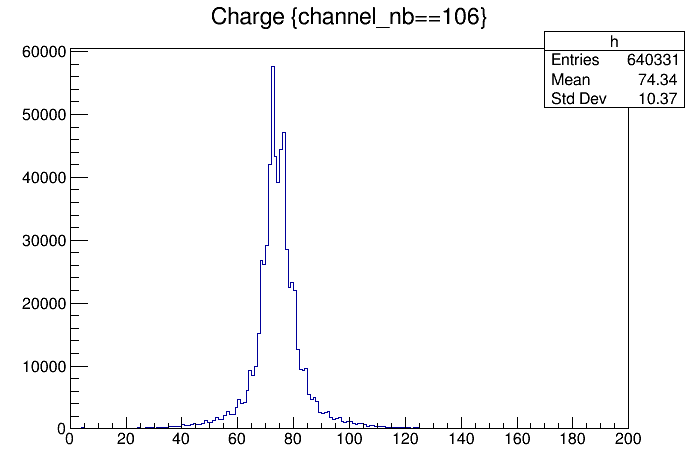}}
     \subfigure[Pedestal width of all channels in high and coarse range]{\includegraphics[width=0.48\textwidth]{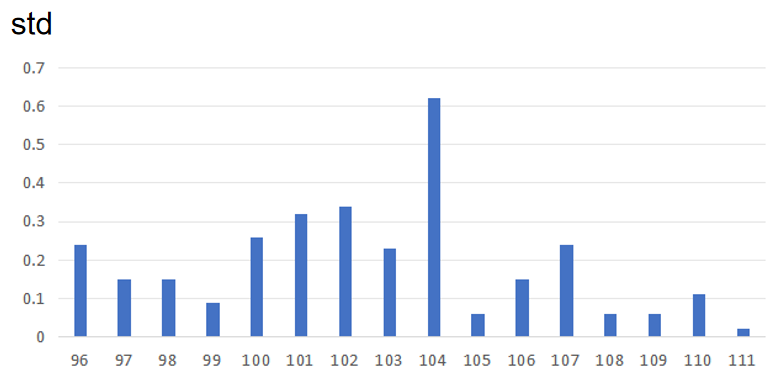}}
    \caption{Pedestal of all used 16 channels}
    \label{fig:l}
\end{figure}

The CATIROC electronics can be run under pedestal (high gain or coarse gain) and physics modes, all the 16 PMTs are connected to channels 96-111. The pedestal mode is run without applying voltage, to get the typical value and its standard deviation of the pedestal, which reflects the noise level of the system. For example, Fig.~\ref{fig:l}a shows the pedestal and standard deviation for channel 106. The standard deviation values of the pedestal are showing in Fig.~\ref{fig:l}b.

\begin{figure}[!ht]
    \centering
    \subfigure[charge spectrum of SPE in ADC]{\includegraphics[width=0.45\textwidth]{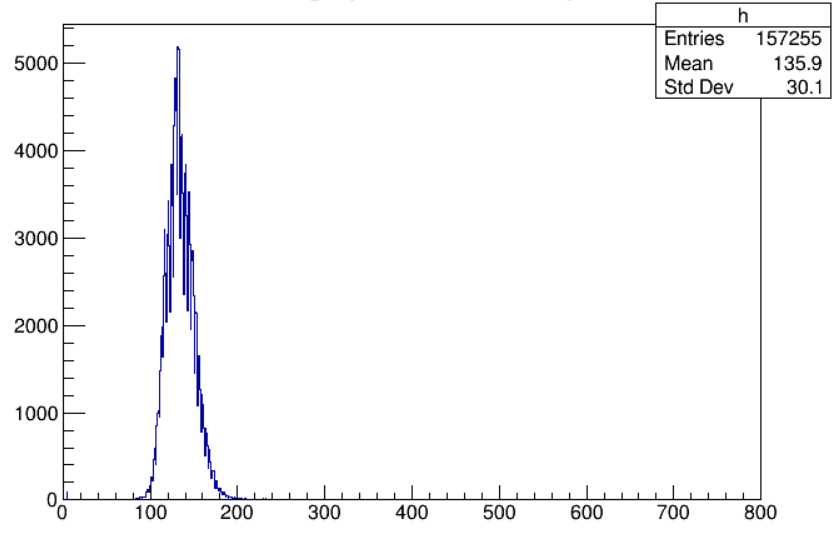}}
     \subfigure[Charge spectrum of SPE in pe.]{\includegraphics[width=0.45\textwidth]{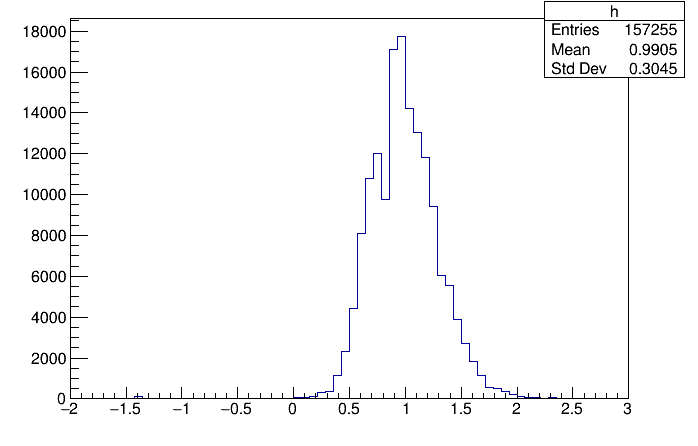}}
    \subfigure[charge spectrum of multi PE in ADC]{\includegraphics[width=0.45\textwidth]{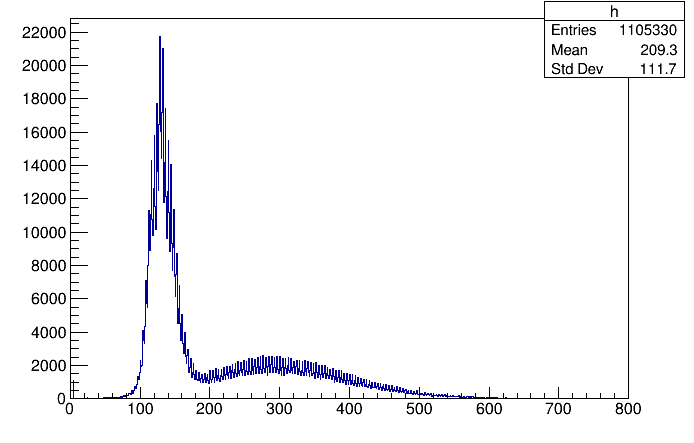}}
     \subfigure[Charge spectrum of multi PE in pe.]{\includegraphics[width=0.45\textwidth]{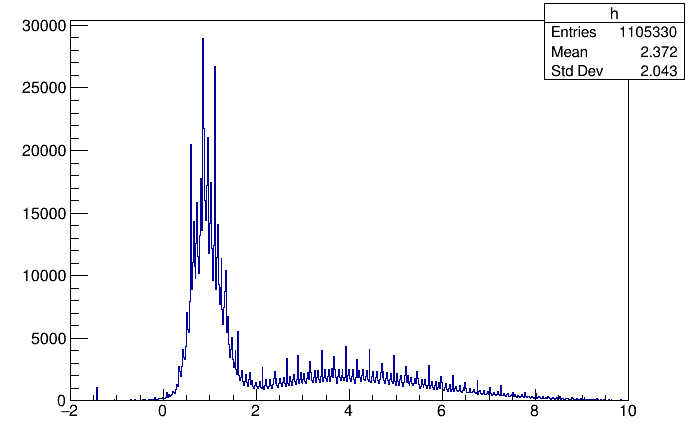}}
    \caption{Measured charge spectrum by Catiroc}
    \label{fig:spe-combine}
\end{figure}

Fig.\,\ref{fig:spe-combine}a and \ref{fig:spe-combine}b respectively shows an example of the CATIROC measured single photon electron spectrum (SPE) in ADCu and coverted to 
p.e.\,with a threshold of around 0.3\,p.e., and the dark count rate of the PMT can also be calculated according to the trigger counts and running time. With the measured SPE spectrum, the PMT gain is re-measured on-site. Fig.\,\ref{fig:spe-combine}c and \ref{fig:spe-combine}d are the charge spectra of multi-photoelectron with the LED on, where we also can find the SPE peak from the triggerless data taking of each PMT channel.
With the taken data, an offline matching algorithm applied to all the 16 PMT channels after sorting all the hits according to hit time (coarse time + fine time) to identify possible events. The trigger matching window is settable and tested with 100\,ns, 200\,ns, 500\,ns, 1000\,ns.


\newpage


\subsection{Design and installation of the prototype}

\begin{figure}[!ht]
        \subfigure[The shell of the prototype]{\includegraphics[width=0.45\textwidth]{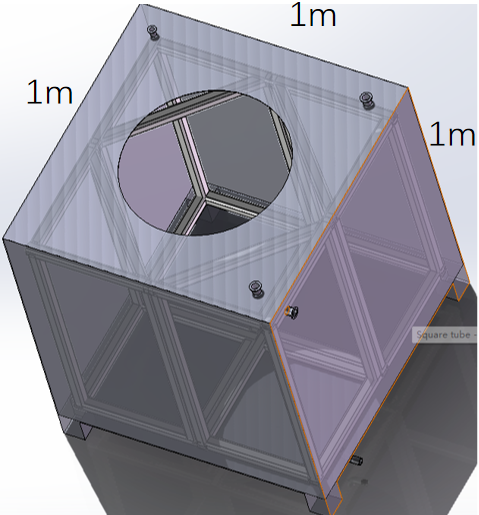}}
     \subfigure[Internal brackets.]{\includegraphics[width=0.47\textwidth]{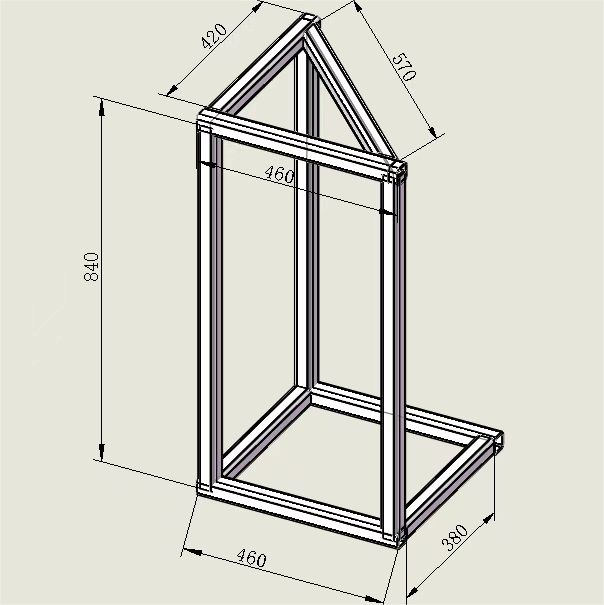}}  
  \caption{ Design drawing of the structure of the water tank prototype. }
\label{fig.a}
  \end{figure}

The shell of the water tank prototype is designed to 1\,m $\times$ 1\,m $\times$ 1\,m by stainless steel (Fig.~\ref{fig.a}a), with a manhole in the middle of the top for installation of internal brackets and PMTs. An water outlet is designed on its bottom and an overflow pipe on the side near to the highest top cover. There are three ports on the top surface as cable feed through. The internal bracket assembled by standard alumina profile is designed for supporting PMTs and LED, and a tyvek of type 1082D with thickness 600\,um is fully covered the internal surface of the tank. A pre-assembly sub-structure (Fig.~\ref{fig.a}b), divided the bracket into four parts, is utilized for the installation, which are pre-assembled externally and then connected internally to form the complete structure. The design schematic diagram of the prototype water tank shell and internal bracket is shown in Fig.~\ref{fig.a}.

\begin{figure}[!ht]
    \centering
    \subfigure[Water tank prototype]{\includegraphics[width=0.45\textwidth]{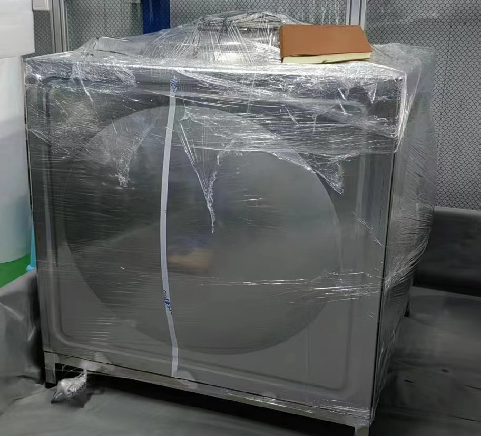}}
    \subfigure[Manhole of the tank]{\includegraphics[width=0.45\textwidth]{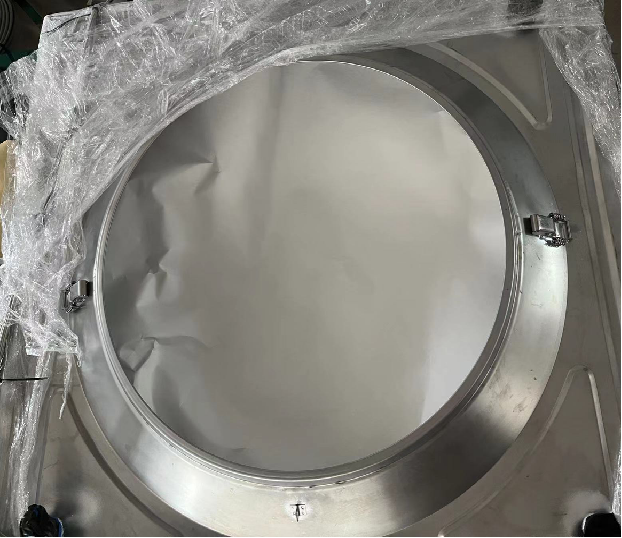}}
    \subfigure[Tyvek, structure, LED and PMT installed]{\includegraphics[width=0.45\textwidth]{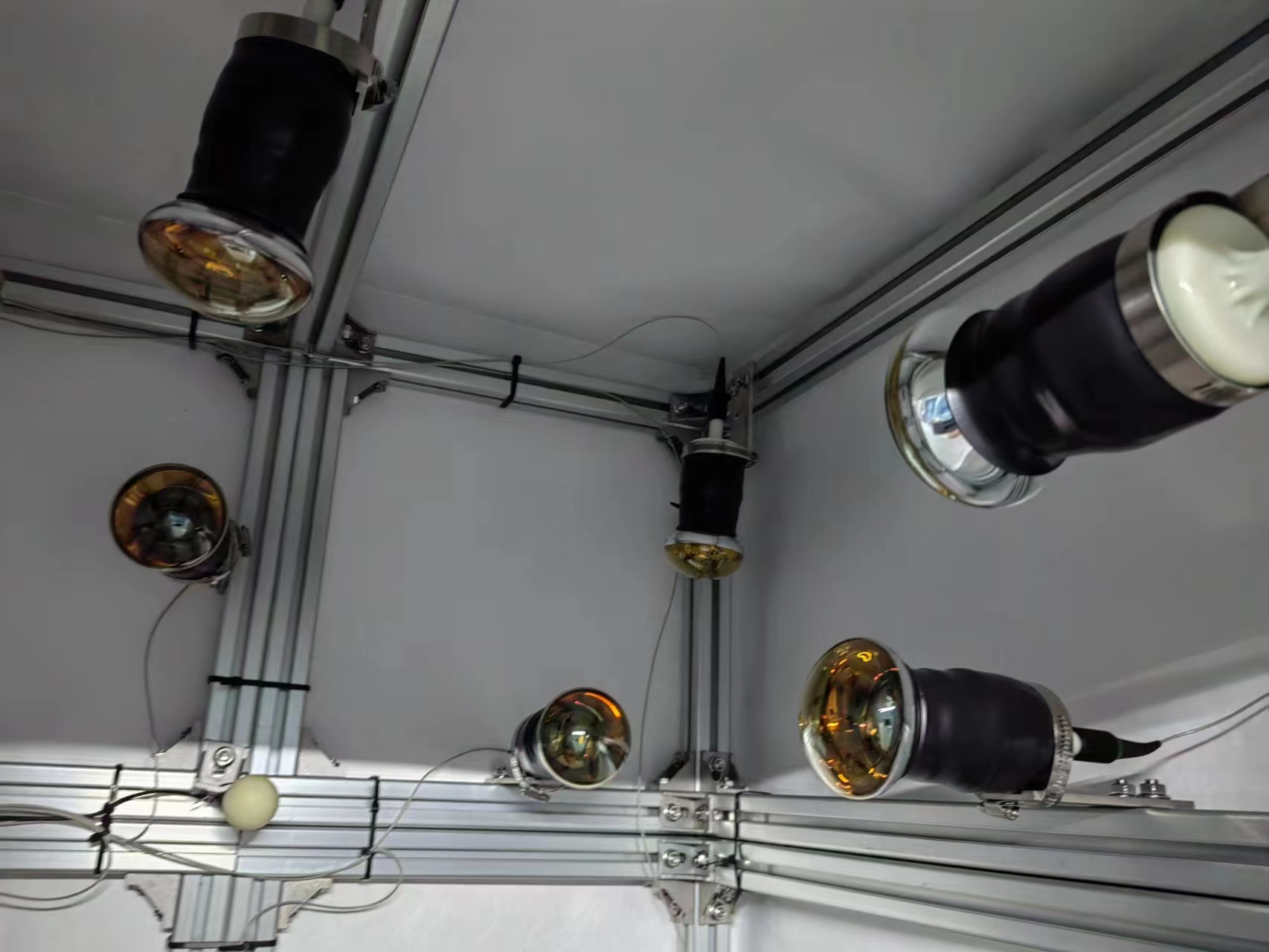}}
     \subfigure[covered tank and the used electronics UWB]{\includegraphics[width=0.45\textwidth]{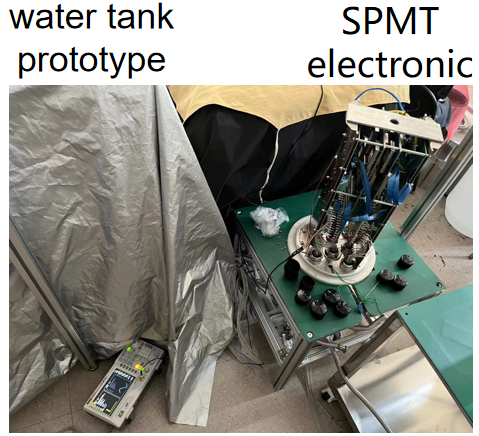}}
     \caption{The water tank prototype system.}
    \label{fig:b}
\end{figure}

 During installation, a pre-cleaning process was carried out by washing each part with pure water and wiping them with alcohol. The weld seams of the shell were coated with epoxy paint. Assemble the bracket and 16 PMTs, and wrap the bracket with Tyvek inside the tank to ensure uniform light distribution inside. The position of SPMTs is evenly distributed (four viewing up on bottom, four viewing side on the corner, four viewing inside located at the middle height, and another four viewing down from the top), and an LED with a diffuser ball is placed in the middle at the bottom for subsequent testing. Then pull the cable out of the cable feed through, cover the opening, seal the cable outlet, and wrap the surface with plastic film (without nitrogen flushing). Finally, cover the water tank prototype with black clothes for light tight. Fig.~\ref{fig:b} shows the internal distribution of PMT installation, Tyvek, and the water tank prototype with electronic.


\newpage

\section{Integration of dry water tank}
\label{1:exp} 

\subsection{PMT characterize in detector}
\label{2:calib}

After the installlation of the 16 PMTs into the water tank, all of them are connected to a single CATIROC of the UWB through a 10\,m cable. Following the design of JUNO UWB, it can only be supplied with the same voltage to all the channels connected to a single CATIROC, which is set at 1200\,V here. In the dark state, data is collected with the dark noise to check the SPE of each PMT. With the conversion relationship between the electronic charge unit (ADCu) of each channel and pico-Column (pc), the gain for the 16 channels was determined, which is basically consistent with the previously measure results. But the channels show a big difference from the same working HV as expected.

\captionsetup[table]{justification=centering}
\captionsetup[table]{skip=1pt}
\begin{table}[!ht]
	\centering
	\caption{Gain of each PMT}
	\begin{minipage}{.5\textwidth}
		\centering
		\begin{tabular}{ccc}
			\hline
			Channel number & Gain@1200V \\
			\hline
			No.96 & 2.8e6 \\
			No.97 & 1.9e6 \\
			No.98 & 3.3e6 \\
    No.99 & 2.7e6 \\ 
			No.100 & 1.8e6 \\
			No.101 & 2.5e6 \\
   No.102 & 2.0e6 \\
   No.103 & 4.4e6\\
   \hline
		\end{tabular}
	\end{minipage}%
	\begin{minipage}{.5\textwidth}
		\centering
		\begin{tabular}{ccc}
			\hline
			Channel number & Gain@1200V \\
			\hline
			No.104 & 2.1e6 \\
			No.105 & 11.0e6 \\
			No.106 & 8.4e6 \\
   No.107 & 2.0e6 \\
    No.108 & 23.0e6 \\
   No.109 & 9.9e6 \\
    No.110 & 2.9e6 \\
   No.111 & 15.0e6 \\
   \hline
		\end{tabular}
	\end{minipage}
 \label{tab.gain}
\end{table}


The PMTs' dark count rate (DCR) was measured using electronics as discussed, obtaining the DCR for each PMT. The DCR of each PMT is as shown in Tab.~\ref{tab.DCR}. It is also showing a big difference suffering from the gain difference and PMT itself.

\captionsetup[table]{justification=centering}
\captionsetup[table]{skip=1pt}
\begin{table}[!ht]
	\centering
	\caption{DCR of each PMT}
	\begin{minipage}{.5\textwidth}
		\centering
		\begin{tabular}{ccc}
			\hline
			SPMT number & DCR/Hz \\
			\hline
			No.96 & 2102 \\
			No.97 & 321 \\
			No.98 & 2434 \\
    No.99 & 977 \\
			No.100 & 91 \\
			No.101 & 2264 \\
   No.102 & 625 \\
   No.103 & 1354\\
   \hline
		\end{tabular}
	\end{minipage}%
	\begin{minipage}{.5\textwidth}
		\centering
		\begin{tabular}{cc}
			\hline
			SPMT number & DCR/Hz \\
			\hline
			No.104 & 56 \\
			No.105 & 1603 \\
			No.106 & 24250 \\
   No.107 & 388 \\
    No.108 & 8162 \\
   No.109 & 16114 \\
    No.110 & 1406 \\
   No.111 & 971 \\
   \hline
		\end{tabular}
	\end{minipage}
 \label{tab.DCR}
\end{table}

\newpage
\subsection{Event assembly with LED}
\label{2:LED}

\begin{figure}[!ht]
    \centering
    \includegraphics[width=1\linewidth]{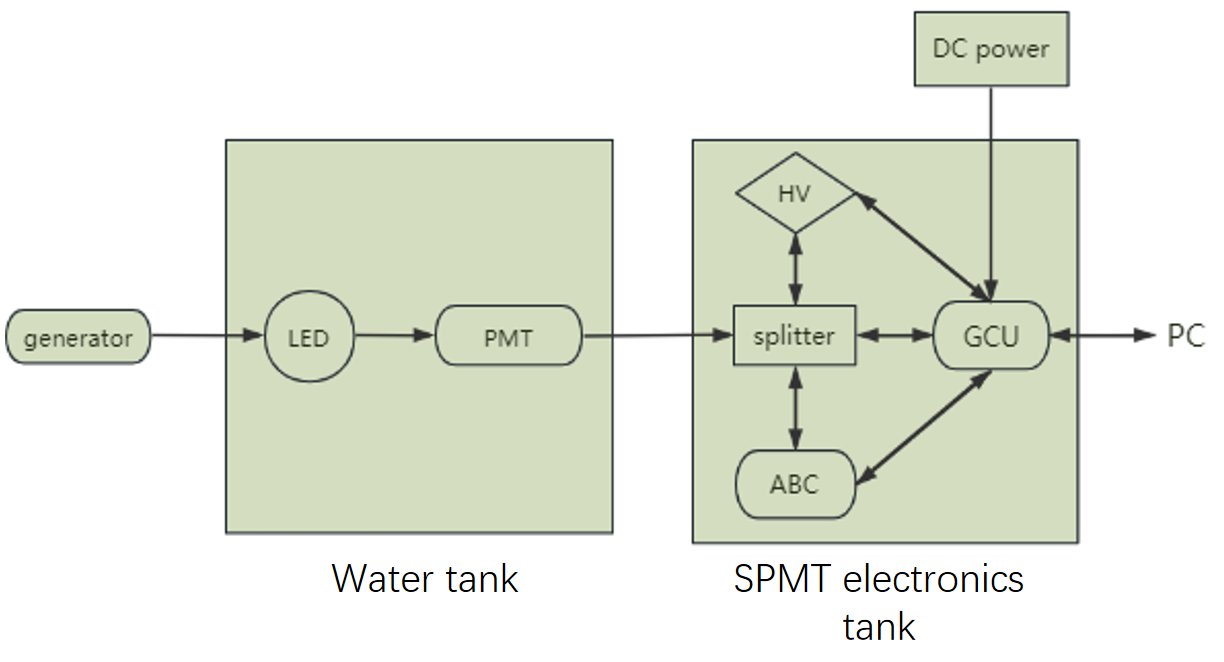}
    \caption{The logical block diagram of the water tank + electronics + LED.}
    \label{fig:x}
\end{figure}

Fig.~\ref{fig:x} shows the logical block diagram of the dry water tank + electronics + LED, which also runs under triggerless mode. 

\begin{figure}[!ht]
                \subfigure[different intensities]{\includegraphics[width=0.45\textwidth]{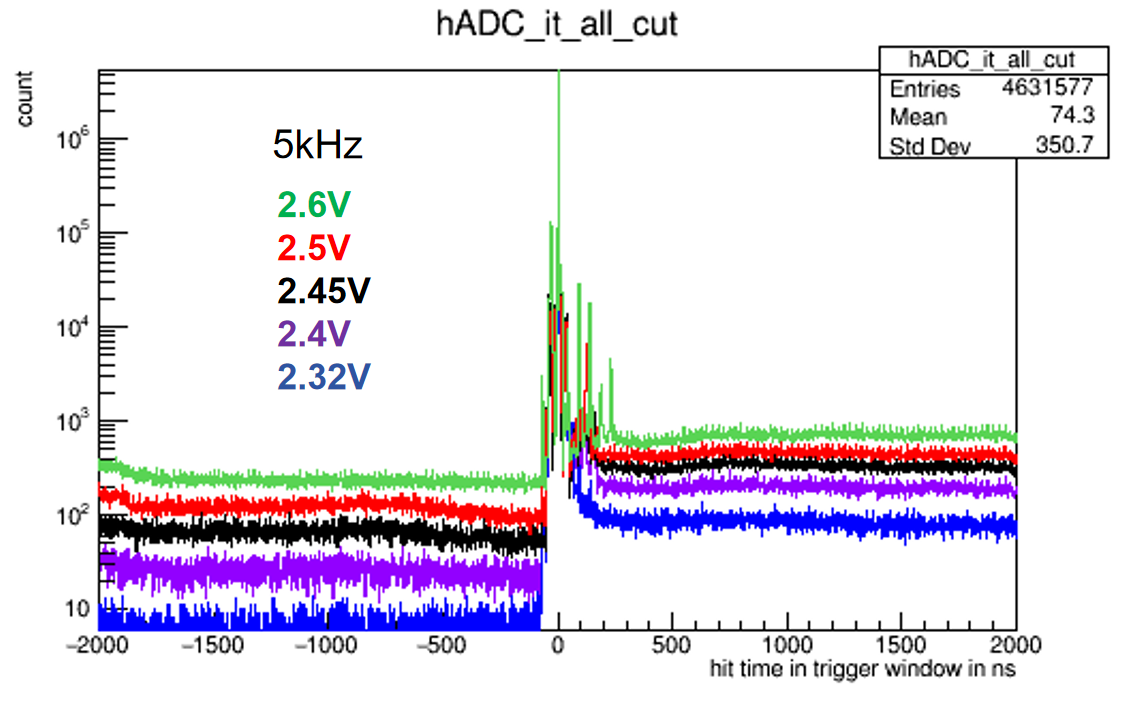}}
                \subfigure[different frequencies]{\includegraphics[width=0.45\textwidth]{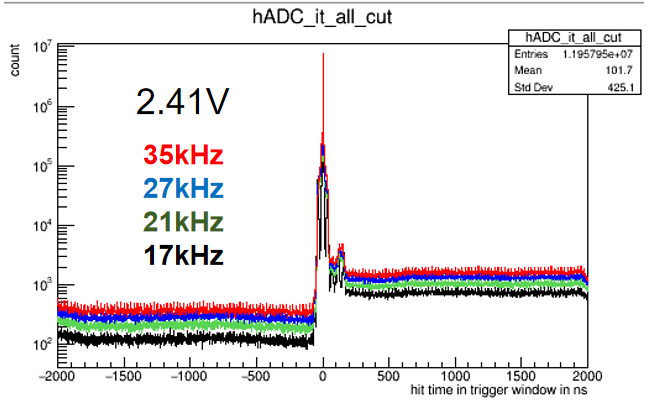}}
  \caption{ Distribution of hit time under different light intensities and different illumination frequencies of the LED. }
		\label{fig：time window}
  \end{figure}

With the taken data, an event window of [-2,+2]\,us among the hits is selected to assembly the event after time sorting all hits: if the time difference among channels hits is within the time window, they are considered as an event. Fig.~\ref{fig：time window} shows the distribution of hit time under different light intensities or different frequencies of the LED. It is showing that the offline sorting and event assembly basically working well.
It can be seen that most of the hits located around 100\,ns, while the random coincidence after the main peak is a little bit higher than that before it.This indicates that there may be some noise in the data or potential electronic crosstalk.

\begin{figure}[!ht]
     \centering
     \includegraphics[width=1\linewidth]{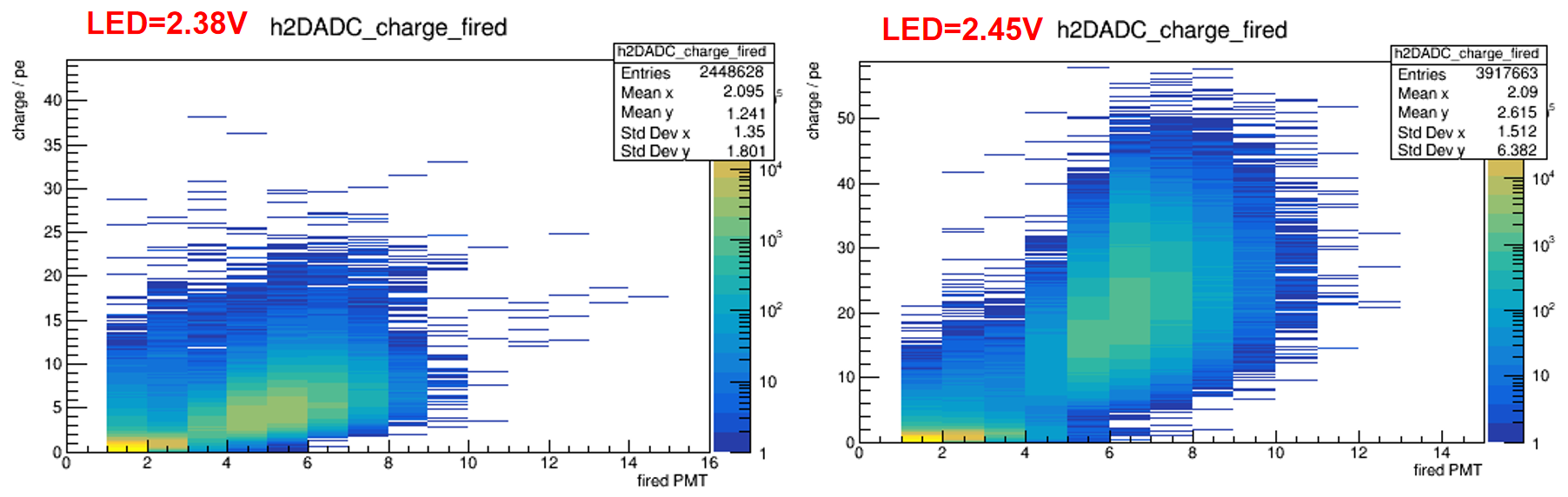}
     \caption{the total PE (SumPE) vs.\,fired PMT of different LED intensities.}
     \label{fig:SumPE vs. firedPMT}
 \end{figure}

 Fig.~\ref{fig:SumPE vs. firedPMT} is a two-dimensional plot showing the total charge (in pe) and the number of fired PMTs of one event. It can be observed that as the light intensity increasing, both the total charge and the number of fired PMTs are increasing, which confirms the sorting and event assembly.
 
\begin{figure}[!ht]
            \subfigure[Different intensities]{\includegraphics[width=0.48\textwidth]{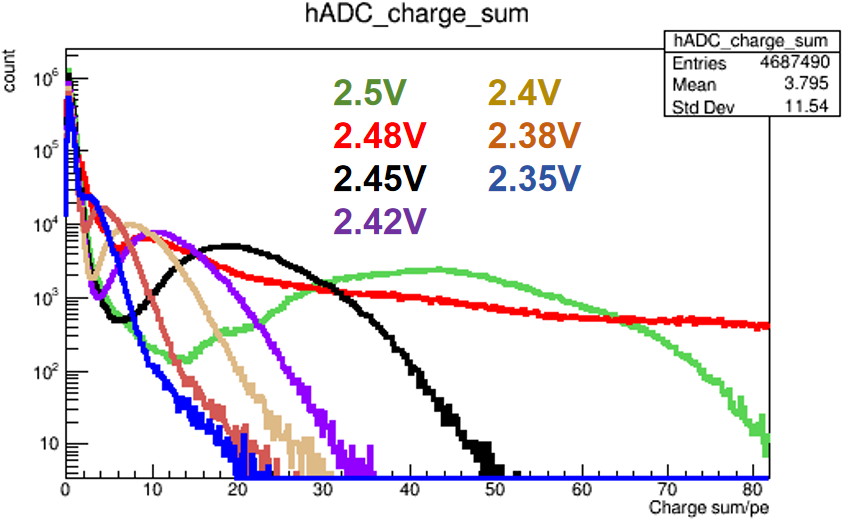}}
            \subfigure[Different frequencies]{\includegraphics[width=0.48\textwidth]{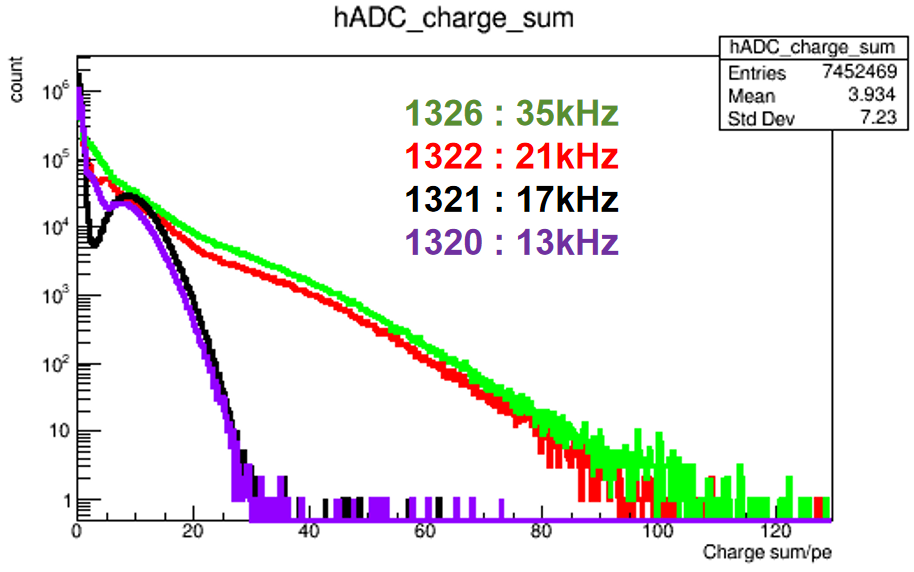}}
  \caption{ Charge spectrum under different light intensities and different illumination frequencies of the LED.}
			\label{fig:charge}
  \end{figure}

Fig.~\ref{fig:charge} shows the charge spectrum under different light intensities when the LED frequency is 5\,kHz, and different frequencies at an intensity of 2.41\,V. It can be observed that when the LED intensity is appropriate, a signal peak is found, but the distribution is wide, and it shifts to the right as the light intensity increases, indicating that the system responds as expected. When the LED frequency is appropriate, the system remains stable when the event rate lower than 17\,kHz, which should be related to the CATIROC features\,\cite{catiroc-10}.

\section{Muon with water filled tank}
\label{1:results}

 \subsection{Muon Rate}
\label{1:rate}

\begin{figure}[!ht]
			\centering	
        \subfigure[Fired PMT versus total charge in PE]{\includegraphics[width=0.48\textwidth]{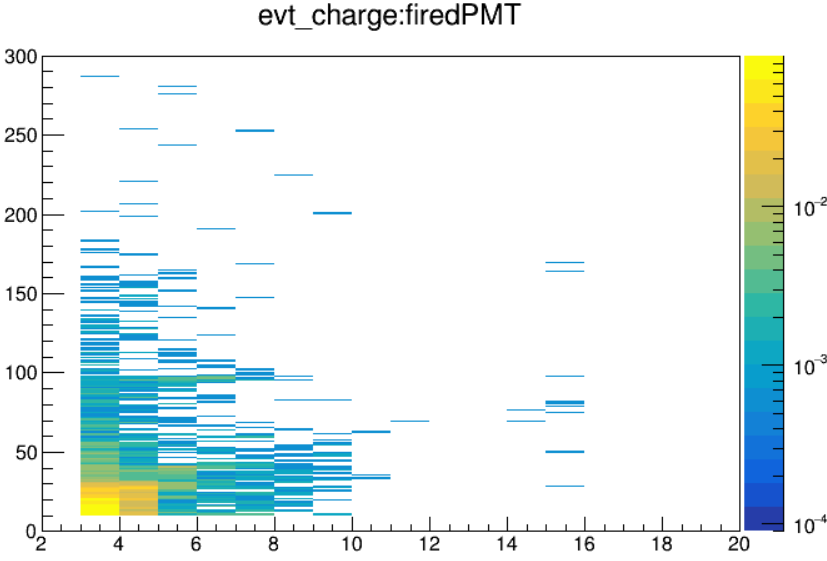}}
        \subfigure[Event rate verses trigger window]{\includegraphics[width=0.48\textwidth]{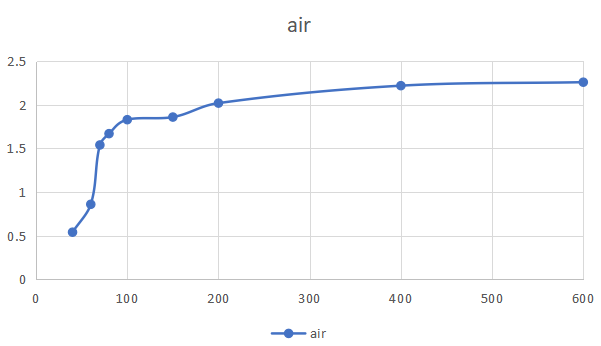}}
           \caption{ System test in air }
			\label{fig:air}
  \end{figure}

The water tank prototype is tested in air without LED before water filling, and a 2-D histogram of fired PMT versus total charge in p.e.\,after event assembly in a window of [-200,200]\,ns is shown in Fig.~\ref{fig:air}. Here a trigger threshold for muon is selected as fired PMT larger or equal to three, and event charge larger and equal to 10\,p.e. A trigger rate verses offline coincidence window also shows that with a longer than 100\,ns window ([-50,50]\,ns) the trigger rate is stable. The random coincidence of 16 PMTs with 5\,kHz dark rate per PMT and 100\,ns window is around 1.83\,Hz. The random coincidence of 16 PMTs, with an average dark rate of 4\,kHz per PMT and a 100\,ns window, is around 1.8\,Hz.

\begin{figure}[!ht]
                \subfigure[Fired PMT of event]{\includegraphics[width=0.48\textwidth]{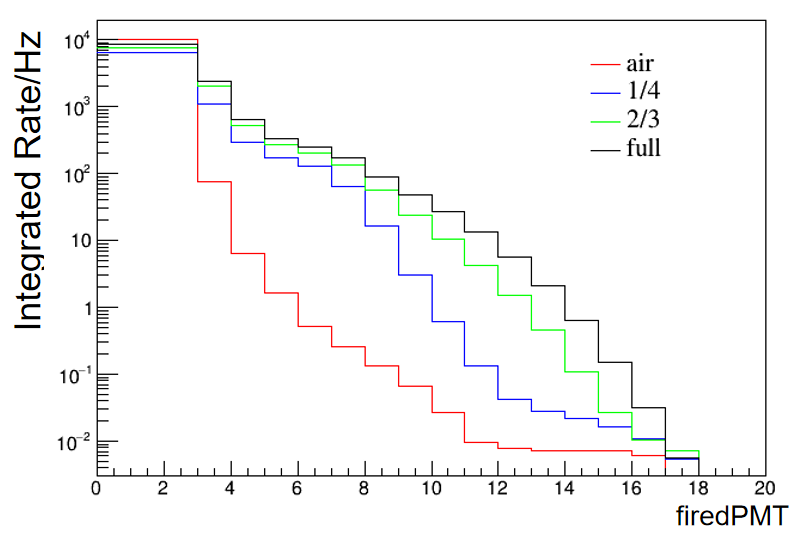}}
                \subfigure[Event charge]{\includegraphics[width=0.48\textwidth]{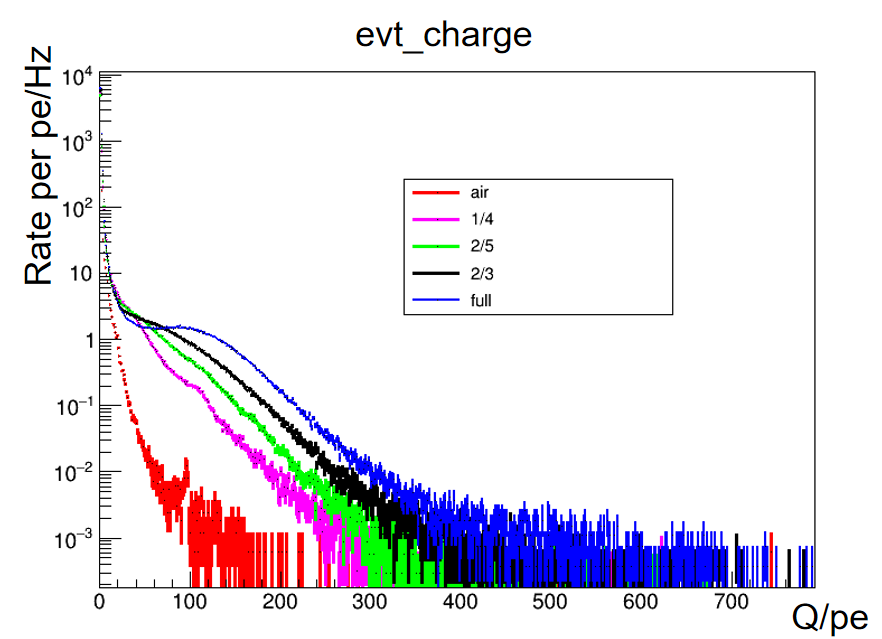}}
  \caption{ Data of different water levels. }
			\label{fig:filling}
  \end{figure}

During the water filling of the prototype, data is taken at several water levels of 1/4, 2/3, full. Fig.~\ref{fig:filling} shows the trigger rate of fired PMTs and event charge in p.e.\,after offline event assembly in a window of [-200,200]\,ns. It can be observed that both the number of fired PMTs and the charge in p.e.\,are increasing when the water level increasing as expected. It can be identified of a hint of a peak around eight of fired PMT or around 100\,p.e.\,on event charge.

\begin{figure}[!ht]
                \subfigure[Fired PMT verses total charge]{\includegraphics[width=0.48\textwidth]{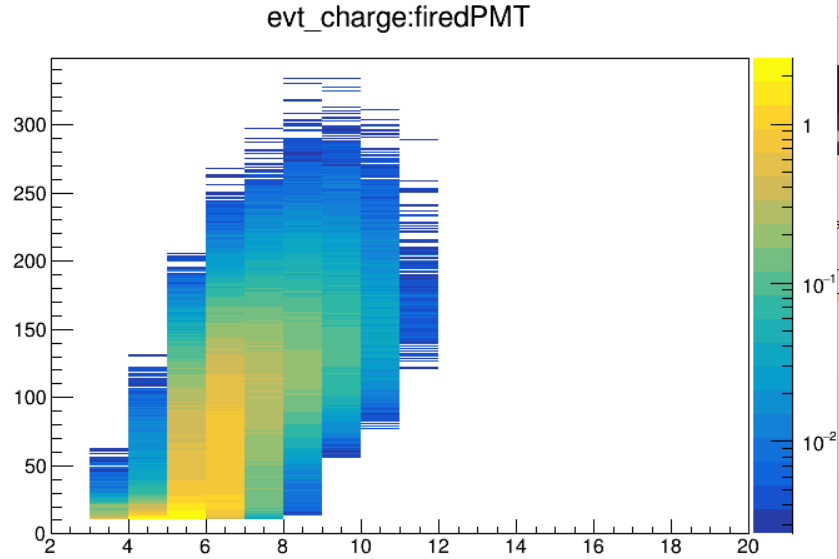}}
                \subfigure[The trigger rate versus coincidence window]{\includegraphics[width=0.48\textwidth]{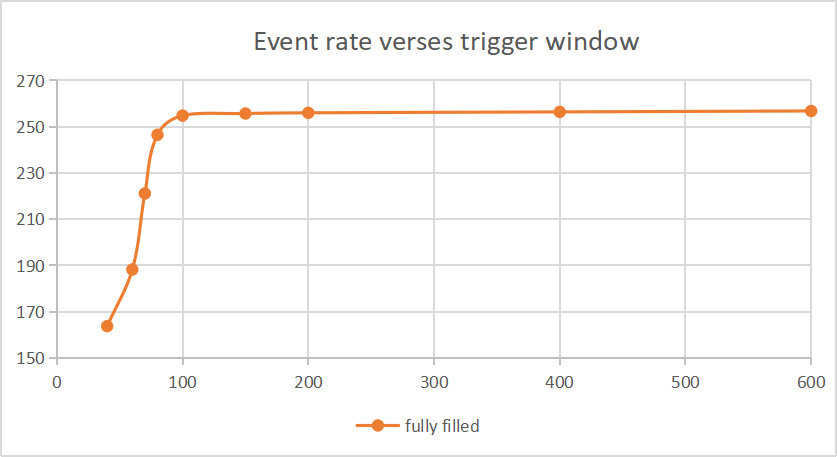}}
         \caption{ Data when water fully filled }
			\label{fig:2d}
  \end{figure}

The 2-D plot of fired PMT versus event charge is further checked for fully filled tank as shown in Fig.~\ref{fig:2d}a. With the features, a threshold of muon identification is proposed as fired PMT equal and larger than 3 (individual fired channel charge equal and larger than 0.3), and event charge in p.e.\,equal and larger than 10. Fig.~\ref{fig:2d}b shows the rate of events exceeding the threshold after water filling using different time windows. Comparing with the rate in air, it can be determined that the time window can be set to 100 ns ([-50,50] ns) after water filling.

 \subsection{Stability}
\label{1:stable}

\begin{figure}[!ht]
     \centering
     \includegraphics[width=0.6\linewidth]{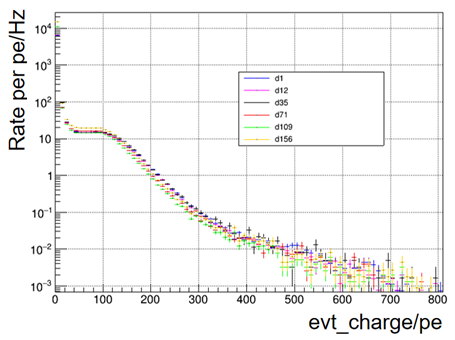}
     \caption{The change of p.e. with time}
     \label{fig:charge vs. time}
 \end{figure}

The change of muon rate versus time is shown in Tab.~\ref{tab.muon rate} after the tank filled in air and sealed there. Moreover, Fig.~\ref{fig:charge vs. time} shows the p.e. spectra versus time. According to the results, it seems for muon detection there is no big decay in 160 days at least.

 \captionsetup[table]{justification=centering}
\captionsetup[table]{skip=1pt}
\begin{table}[!ht]
	\centering
	\setlength{\tabcolsep}{4.5mm}{
		\caption {The change of muon rate with time.}
		\begin{tabular}{cc}
			\hline
			Time(day) & Rate(Hz) \\
			\hline
			1 &  255\\
			7 &  255 \\
			12 &  256\\
           23 & 254 \\
            38 & 259 \\
            65 & 263 \\
            72 & 255 \\
            87 & 251 \\  
            109 & 251 \\
            156 & 246 \\
    \hline
			\label{tab.muon rate}	
	\end{tabular}}
\end{table}

\section{Simulation}
\label{1:sim}

Using Geant4 to simulate muons hitting the small water tank prrototype, with the size of the tank set as 1m*1m*1m. Cosmic rays are sampled proportionally on five faces of the tank except the bottom face. The number of muons per run is fixed, and the results are normalized . The quantum efficiency is set to the quantum efficiency corresponding to different wavelengths actually measured in the experiment with 3-inch PMTs. By changing the attenuation length of water and the reflectivity inside the small water tank and comparing with experimental data, the actual parameters inside the tank can be obtained.

\subsection{Light Yield}
\label{1:yield}

\begin{figure}[!ht]
     \centering
     \includegraphics[width=0.6\linewidth]{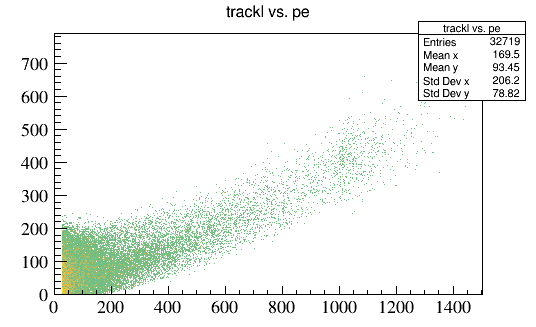}
     \caption{Track length (X) in mm vs. p.e.(Y)}
     \label{fig:tracklpe}
 \end{figure}

Checking the impact of attenuation length, quantum efficiency (QE), and reflectivity on light yield. Reflecting the light yield level through the slope of the two-dimensional graph of track length and p.e., as shown in Fig.~\ref{fig:tracklpe}.

\begin{figure}[!ht]
        \centering
            \subfigure[Tyvek reflectivity]{\includegraphics[width=0.6\textwidth]{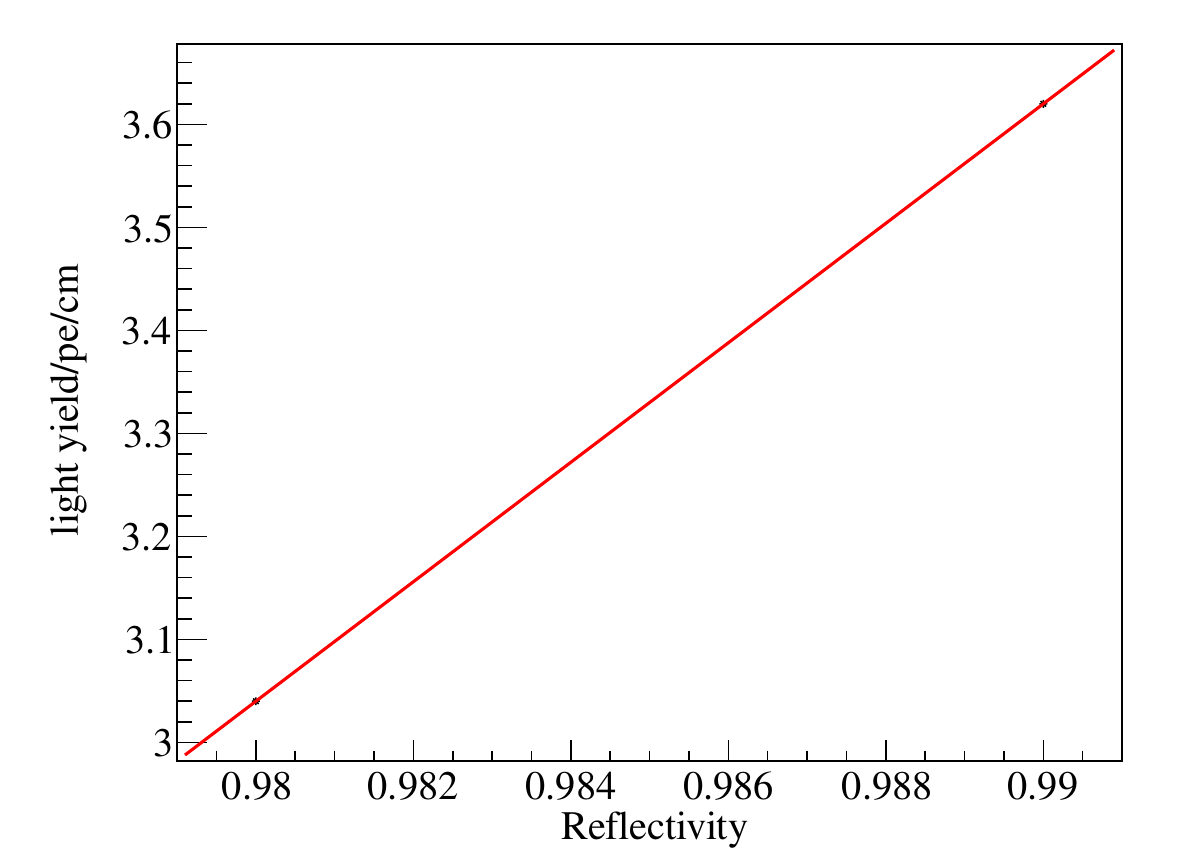}}
            \subfigure[PMT quantum efficiency (QE)]{\includegraphics[width=0.48\textwidth]{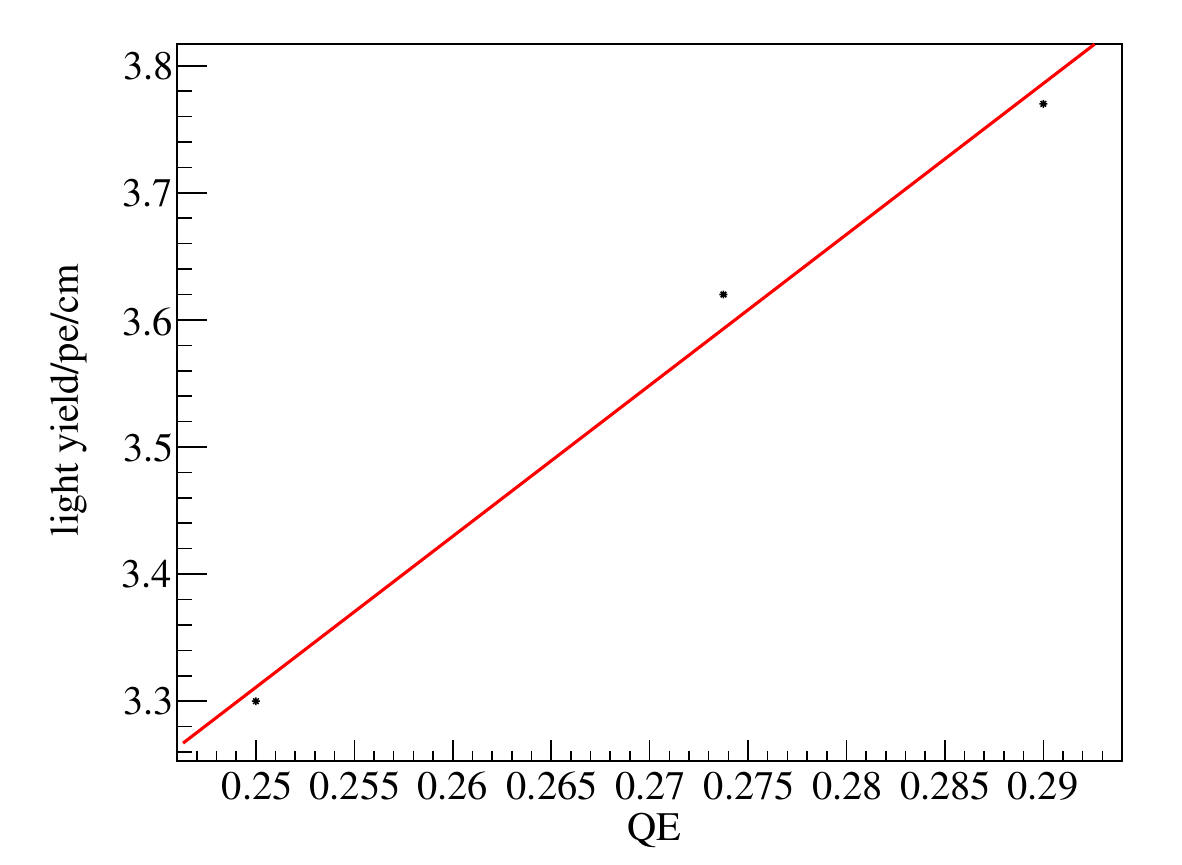}}
            \subfigure[Attenuation length]{\includegraphics[width=0.48\textwidth]{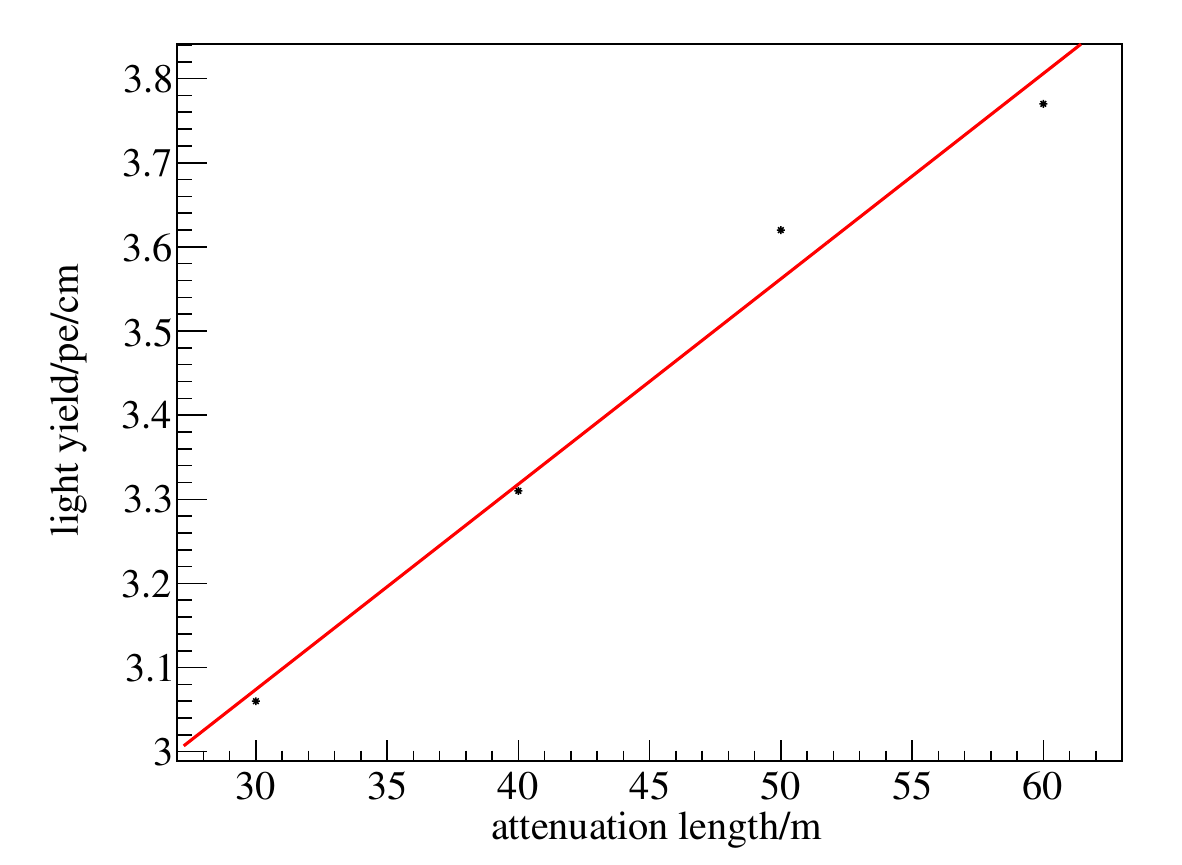}}
\caption{Impact of key parameters to the light yield of the prototype in PE/cm }
			\label{fig:yield}
  \end{figure}

Adjusting the reflectivity from 0.98 to 0.99 with a attenuation length of 50m and QE is set to 27\% based on the experimentally measured quantum efficiencies for different wavelengths, which is also very close to the average value provided by the HZC company. The light yield variation is shown in Fig.~\ref{fig:yield}a, from 3.0\,p.e./cm to 3.6\,p.e./cm. Adjusting the QE from 25\% to 29\% with a reflectivity of 0.99 and attenuation length of 50\,m, the light yield variation is shown in Fig.~\ref{fig:yield}b, from 3.3\,p.e./cm to 3.8\,p.e./cm. Adjusting the attenuation length from 30\,m to 60\,m with a QE of 27\% and a reflectivity of 0.99, the light yield variation is shown in Fig.~\ref{fig:yield}c, from 3.1\,p.e./cm to 3.8\,p.e./cm.


\begin{figure}[!ht]
\centering
\subfigure[Data vs.\,attenuation lengths]{\includegraphics[width=0.75\textwidth]{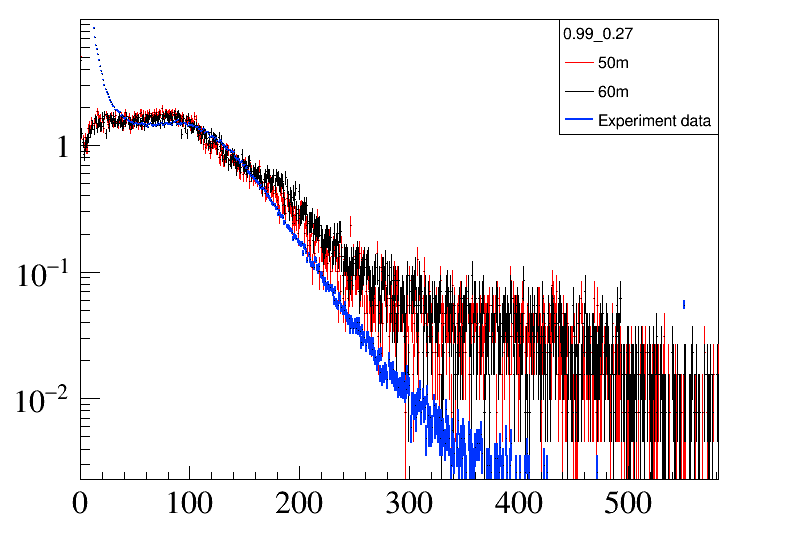}}
\caption{Charge spectrum of Data versus simulation}
\label{fig:simu}
\end{figure}

 According to the simulation, the reflectivity of around 0.99 seems more reasonable, with a quantum efficiency to 27\%, comparing the charge spectra with different attenuation lengths with the experimental charge spectrum as shown in Fig.~\ref{fig:simu}, and it was found that the attenuation length around 60\,meters is close to the experimental data.
 While the difference between data and simulation when the charge is higher than 170\,pe, it is mainly from the electronics effect.The discrepancy between data and simulation in the low-energy region is due to the presence of natural radioactivity and noise in the experimental data.





\section{Summary}
\label{1:summary}

With the small and non-cyclic water tank prototype with 16 3-inch PMTs, the calibration of 3-inch PMTs, JUNO-SPMT electronics used for a water Cerenkov detector, the DAQ and offline software trigger are integrated successfully. The muon rate of the prototype water tank on ground laboratory is measured according to the fired PMT and summed charge under air, during filling or stable running phase. The detector shows a good stability in a five months running without cycling of the water under normal environment 20 to 30 centi degree, which provide a good reference for the future running of JUNO-TAO VETO detector. A simulation also confirms the understanding and feature of the detector.

\acknowledgments
This work was supported partially by the National Natural Science Foundation of China (Grant No. 11875282 and 12022505), the Strategic Priority Research Program of the Chinese Academy of Sciences (Grant No. XDA10011200), and the Youth Innovation Promotion Association of CAS. The authors are most grateful for the support of JUNO SPMT group.

 \bibliographystyle{JHEP}
 \bibliography{biblio.bib}

\providecommand{\href}[2]{#2}\begingroup\raggedright\begin{thebibliography}{10}

\bibitem{2020arXiv200508745J}
{JUNO Collaboration}, A.~{Abusleme}, T.~{Adam}, S.~{Ahmad}, S.~{Aiello}, M.~{Akram} et~al., \emph{{TAO Conceptual Design Report}}, \href{https://doi.org/10.48550/arXiv.2005.08745}{\emph{arXiv e-prints} (2020) arXiv:2005.08745} [\href{https://arxiv.org/abs/2005.08745}{{\ttfamily 2005.08745}}].

\bibitem{Steiger2022TAOTheTA}
H.T.J.~Steiger, \emph{Tao—the taishan antineutrino observatory}, {\emph{Instruments} (2022) }.

\bibitem{JUNO-2022103927}
J.~Collaboration, \emph{Juno physics and detector}, \href{https://doi.org/https://doi.org/10.1016/j.ppnp.2021.103927}{\emph{Progress in Particle and Nuclear Physics} {\bfseries 123} (2022) 103927}.

\bibitem{Li_2022}
R.~Li, G.~Cao, J.~Cao, Y.~Li, Y.~Wang, Z.~Wang et~al., \emph{Detector optimization to reduce the cosmogenic neutron backgrounds in the tao experiment}, \href{https://doi.org/10.1088/1748-0221/17/09/P09024}{\emph{Journal of Instrumentation} {\bfseries 17} (2022) P09024}.

\bibitem{li2022ambientneutronmeasurementtaishan}
R.~Li, Y.~Li, Z.~Wang, Q.~Li, L.~Zhan and J.~Cao, \emph{Ambient neutron measurement at taishan antineutrino observatory},  2022.

\bibitem{limin-compact-plastic}
L.~Min et~al., \emph{Performance of compact plastic scintillator strips with wavelength shifting fibers using a photomultiplier tube or silicon photomultiplier readout}, \href{https://doi.org/https://doi.org/10.1007/s41365-023-01175-6}{\emph{NUCL SCI TECH} {\bfseries 34} (2023) }.

\bibitem{Luo2023DesignOO}
L.~Guang et~al., \emph{Design optimization of plastic scintillators with wavelength-shifting fibers and silicon photomultiplier readouts in the top veto tracker of the juno-tao experiment}, \href{https://doi.org/10.1007/s41365-023-01263-7}{\emph{NUCL SCI TECH} {\bfseries 34} (2023) }.

\bibitem{auger-prototype-ABRAHAM200450}
J.~Abraham, M.~Aglietta, I.~Aguirre, M.~Albrow, D.~Allard, I.~Allekotte et~al., \emph{Properties and performance of the prototype instrument for the pierre auger observatory}, \href{https://doi.org/https://doi.org/10.1016/j.nima.2003.12.012}{\emph{Nuclear Instruments and Methods in Physics Research Section A: Accelerators, Spectrometers, Detectors and Associated Equipment} {\bfseries 523} (2004) 50}.

\bibitem{auger-2015172}
T.P.A.~Collaboration, \emph{The pierre auger cosmic ray observatory}, \href{https://doi.org/https://doi.org/10.1016/j.nima.2015.06.058}{\emph{Nuclear Instruments and Methods in Physics Research Section A: Accelerators, Spectrometers, Detectors and Associated Equipment} {\bfseries 798} (2015) 172}.

\bibitem{water-prototype-dyb-YU201226}
Z.~Yu, H.~Lu, C.~Yang, L.~Wang, M.~Guan, J.~Liu et~al., \emph{Study of a prototype water cherenkov detector for the daya bay neutrino experiment}, \href{https://doi.org/https://doi.org/10.1016/j.nima.2012.04.050}{\emph{Nuclear Instruments and Methods in Physics Research Section A: Accelerators, Spectrometers, Detectors and Associated Equipment} {\bfseries 682} (2012) 26}.

\bibitem{catiroc-10}
S.~Conforti, A.~Cabrera, C.~De~La~Taille, F.~Dulucq, M.~Grassi, G.~Martin-Chassard et~al., \emph{Catiroc, a multichannel front-end asic to read out the 3{\textacutedbl} pmts (spmt) system of the juno experiment},  in \emph{Proceedings of International Conference on Technology and Instrumentation in Particle Physics 2017}, Z.-A.~Liu, ed., (Singapore), pp.~168--172, Springer Singapore, 2018.

\bibitem{Blin_2017}
S.~Blin, S.~Callier, S.C.D.~Lorenzo, F.~Dulucq, C.D.L.~Taille, G.~Martin-Chassard et~al., \emph{Performance of catiroc: Asic for smart readout of large photomultiplier arrays}, \href{https://doi.org/10.1088/1748-0221/12/03/C03041}{\emph{Journal of Instrumentation} {\bfseries 12} (2017) C03041}.

\bibitem{JUNO-double-miao}
M.~He, \emph{Double calorimetry system in juno}, \href{https://doi.org/10.1007/s41605-017-0022-2}{\emph{Radiat Detect Technol Methods} {\bfseries 1} (2017) }.

\bibitem{linan2021165347-3inch}
N.~Li et~al., \emph{Characterization of 3-inch photomultiplier tubes for the juno central detector}, \href{https://doi.org/https://doi.org/10.1007/s41605-018-0085-8}{\emph{Radiat Detect Technol Methods} {\bfseries 3} (2019) }.

\bibitem{CAO2021165347-3inch}
C.~Cao, J.~Xu, M.~He, A.~Abusleme, M.~Bongrand, C.~Bordereau et~al., \emph{Mass production and characterization of 3-inch pmts for the juno experiment}, \href{https://doi.org/https://doi.org/10.1016/j.nima.2021.165347}{\emph{Nuclear Instruments and Methods in Physics Research Section A: Accelerators, Spectrometers, Detectors and Associated Equipment} {\bfseries 1005} (2021) 165347}.

\bibitem{diru2021165347-3inch}
D.~Wu et~al., \emph{Study of the front-end signal for the 3-inch pmts instrumentation in juno}, \href{https://doi.org/https://doi.org/10.1007/s41605-022-00324-6}{\emph{Radiat Detect Technol Methods} {\bfseries 6} (2022) 349–360}.

\bibitem{DT5751}
C.A.E.N., \emph{{DT5751}},  2024.

\bibitem{JUNO20inchPMT}
A.~Adam, T.~Ahmad et~al., \emph{{ Mass testing and characterization of 20-inch PMTs for JUNO}}, \href{https://doi.org/10.1140/epjc/s10052-022-11002-8}{\emph{Eur. Phys. J. C} {\bfseries 82} (2022) 1168}.

\bibitem{juno-double-calorimetry-miao}
H.M.~on~behalf of~the JUNO~collaboration, \emph{Double calorimetry system in juno}, \href{https://doi.org/https://doi.org/10.1007/s41605-017-0022-2}{\emph{Radiat Detect Technol Methods} {\bfseries 1} (2017) 21}.

\bibitem{JUNO3inchPMT}
C.~Cao et~al., \emph{{Mass production and characterization of 3-inch PMTs for the JUNO experiment}}, \href{https://doi.org/10.1016/j.nima.2021.165347}{\emph{Nucl. Instrum. Meth. A} {\bfseries 1005} (2021) 165347} [\href{https://arxiv.org/abs/2102.11538}{{\ttfamily 2102.11538}}].

\bibitem{10.1007/978-981-13-1313-4_34}
S.~Conforti, A.~Cabrera, C.~De~La~Taille, F.~Dulucq, M.~Grassi, G.~Martin-Chassard et~al., \emph{Catiroc, a multichannel front-end asic to read out the 3{\textacutedbl} pmts (spmt) system of the juno experiment},  in \emph{Proceedings of International Conference on Technology and Instrumentation in Particle Physics 2017}, Z.-A.~Liu, ed., (Singapore), pp.~168--172, Springer Singapore, 2018.

\bibitem{Signal-SPMT-diru}
D.~Wu, J.~Xu, M.~He, Z.~Wang and Z.~Chu, \emph{Study of the front-end signal for the 3-inch pmts instrumentation in juno}, \href{https://doi.org/https://doi.org/10.1007/s41605-022-00324-6}{\emph{Radiat Detect Technol Methods} {\bfseries 6} (2022) 349–360}.

\end{thebibliography}\endgroup
\end{document}